\documentclass{egpubl}
 
\JournalSubmission    
\usepackage[T1]{fontenc}
\usepackage{dfadobe} 
\usepackage{array} 

\usepackage{cite}  
\BibtexOrBiblatex

\electronicVersion
\PrintedOrElectronic
\ifpdf \usepackage[pdftex]{graphicx} \pdfcompresslevel=9
\else \usepackage[dvips]{graphicx} \fi

\usepackage{egweblnk} 

\usepackage{amsfonts}
\usepackage{amsmath}

\title{Real-time and Controllable Reactive Motion Synthesis via Intention Guidance}

\author[X. Zhang \& Z. Chang \& Q. Men \& H. P. H. Shum]
{\parbox{\textwidth}{\centering Xiaotang Zhang$^{1}$, Ziyi Chang$^{1}$, Qianhui Men$^{2}$, and Hubert P. H. Shum$^{1\dagger}$}
        \\
{\parbox{\textwidth}{\centering $^1$Durham University, United Kingdom\\
        \{xiaotang.zhang, ziyi.chang, hubert.shum\}@durham.ac.uk \\
         $^2$University of Bristol, United Kingdom\\
         qianhui.men@bristol.ac.uk\\
         $\dagger$ Corresponding author
       }
}
}
\begin{document}


\teaser{
 \includegraphics[width=0.9\linewidth]{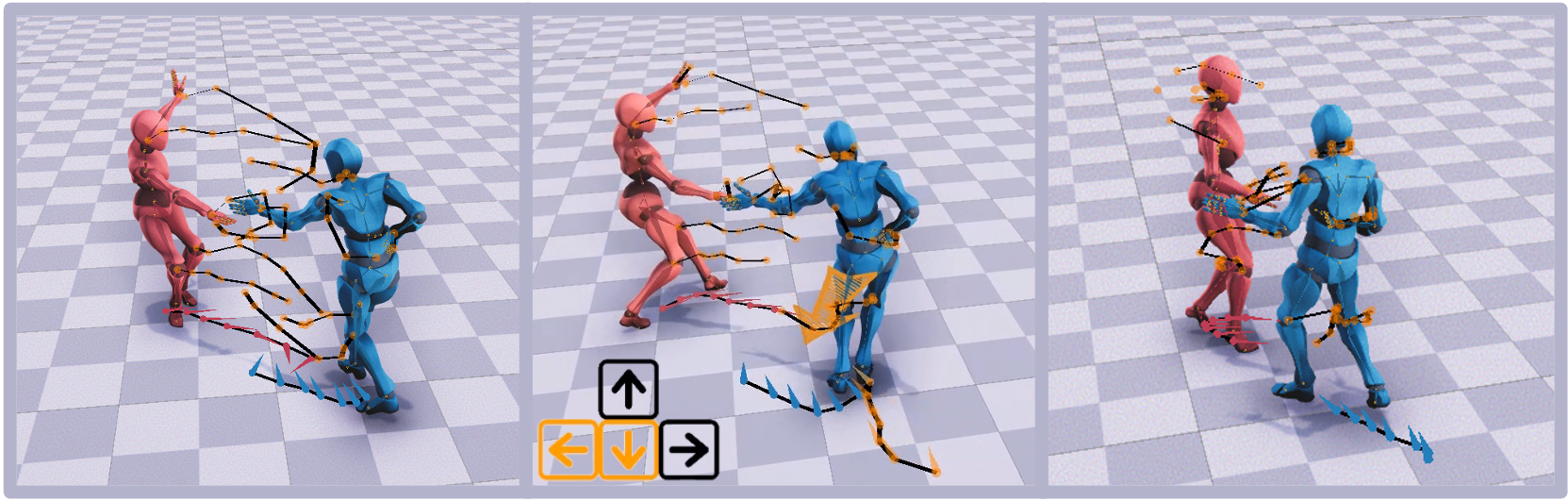}
 \centering
  \caption{Our model generates instant reactive movements for the red character based on anticipated intentions. The user controls the movement direction of the input blue character, and our system dynamically generates corresponding motions for both characters in response.}
\label{Teaser}
}

\maketitle

\begin{abstract}
We propose a real-time method for reactive motion synthesis based on the known trajectory of input character, predicting instant reactions using only historical, user-controlled motions. Our method handles the uncertainty of future movements by introducing an intention predictor, which forecasts key joint intentions to make pose prediction more deterministic from the historical interaction. The intention is later encoded into the latent space of its reactive motion, matched with a codebook which represents mappings between input and output. It samples a categorical distribution for pose generation and strengthens model robustness through adversarial training. Unlike previous offline approaches, the system can recursively generate intentions and reactive motions using feedback from earlier steps, enabling real-time, long-term realistic interactive synthesis. Both quantitative and qualitative experiments show our approach outperforms other matching-based motion synthesis approaches, delivering superior stability and generalizability. In our method, user can also actively influence the outcome by controlling the moving directions, creating a personalized interaction path that deviates from predefined trajectories.

\keywords{animation systems, human simulation, motion capture}
\ccsdesc[500]{Computing methodologies~Motion capture}
\ccsdesc[500]{Computing methodologies~Motion processing}
\printccsdesc
\end{abstract}

\section{Introduction}
Synthesizing motions that are reactive to the other one's movements has been an important task for various applications such as robotics \cite{yoon2019robots} and social analysis \cite{tanke2023social}. By observing the movement history of the other one, reactive motion synthesis aims to obtain interactive motions that respond to the observation. Despite its high demand in various fields, creating reactive motions is usually time-consuming and labor-intensive because of human interactions' diverse and complex dynamics \cite{tanaka2023role,chopin2023interaction}.

Previous approaches for synthesizing reactive motions mainly focus on offline applications. These methods usually require full knowledge of one's movements to generate the corresponding reactive motions \cite{ghosh2024remos,xu2024regennet}. However, their dependence on a full sequence of observed motions is limited for real-world applications because future movements are usually not observable.

We focus on single-character reactive motion synthesis based on the past motion trajectory of input character. Unlike previous methods \cite{ghosh2024remos,xu2024regennet} that generate the reactive motion in offline manner where future motions have already been defined and provided, we attempt to predict the reaction frame-by-frame only based on observed historical motions and also user's control on the input character if provided. As the future of the conditioned motion is unknown and non-deterministic, it requires the neural network to possess real-time and rapid generation capabilities to avoid unrealistic interaction performance in long term.
One possible solution is Motion Matching \cite{gdcvault2024motionmatching,holden2020learned}, which searches for the optimal next pose in a database based on carefully designed query features, making it suitable for real-time inference and instant animation. However, it lacks generalizability and requires the database to encompass all possible interactions when applied to our reactive motion synthesis task.

To achieve instant and high-quality reaction synthesis, rather than directly generating future poses from history states, we introduce an intention predictor that first forecasts the intention of key joints in the input character. Using these auxiliary intention features, we transform the inherently uncertain future pose prediction into a more stable process. Furthermore, inspired by \cite{starke2024categorical}, our model learns a codebook that maps input motion to an appropriate reactive output, avoiding wrong blending artifacts that may appear in motion prediction network. We also introduce adversarial training to promote the learning of a more diverse and effective latent space, leading to improved codebook matching accuracy.

We conduct extensive experiments to quantitatively assess our model's performance on a large-scale dancing dataset, evaluating it across multiple criteria. The results consistently show that our model outperforms other matching-based motion synthesis approaches, achieving superior accuracy and quality in generating reactive motions. To further evaluate its robustness, we introduce variations by modifying input motions based on user-controlled keyboard signals, simulating a range of dynamic conditions (examples given in Figure \ref{Teaser}). In these tests, our model demonstrates strong generalizability, effectively handling diverse input variations while maintaining stable predictions. This stability and adaptability surpass the performance of Motion Matching, highlighting the model’s ability to produce reliable, high-quality outputs even under challenging conditions.

Our contributions can be summarized as:
\begin{itemize}
    \item \textbf{Guidance} - 
    A reactive motion synthesis system guided by input character's intended motion anticipated from historical interactions.
    \item \textbf{Real time} - 
    The system enables online synthesis of human interaction, supporting a continuous loop of the intention input, processing, and feedback of the reactive motion.
    \item \textbf{Controllable} - 
    The system allows users to generate controllable interaction by controlling the footpath to dynamically adjust motion directions of the input character.
\end{itemize}

\begin{figure*}[tbp]
  \centering
  \includegraphics[width=\linewidth]{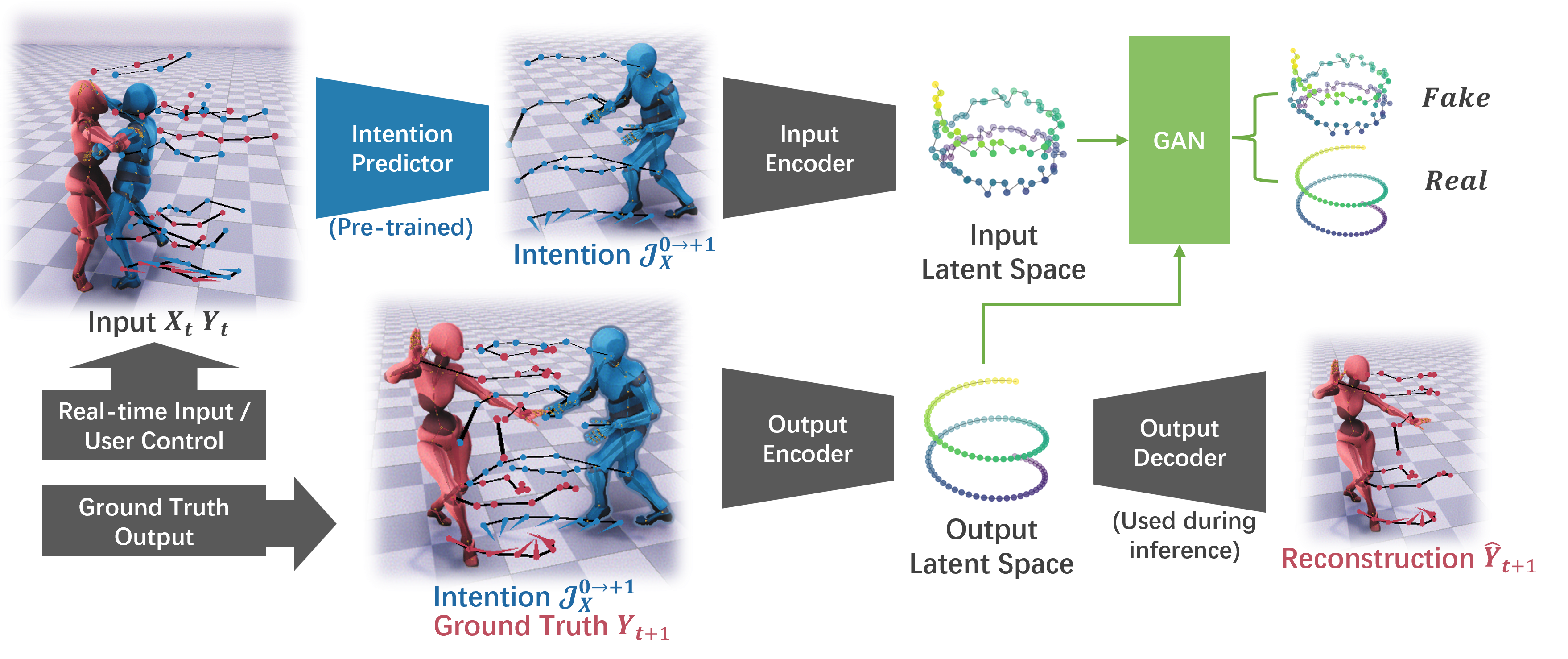}
  \caption{The overall pipeline of our intention-aware reactive motion synthesis system.}
  \label{Architecture}
\end{figure*}

\begin{figure}[tbp]
  \centering
  \includegraphics[width=\linewidth]{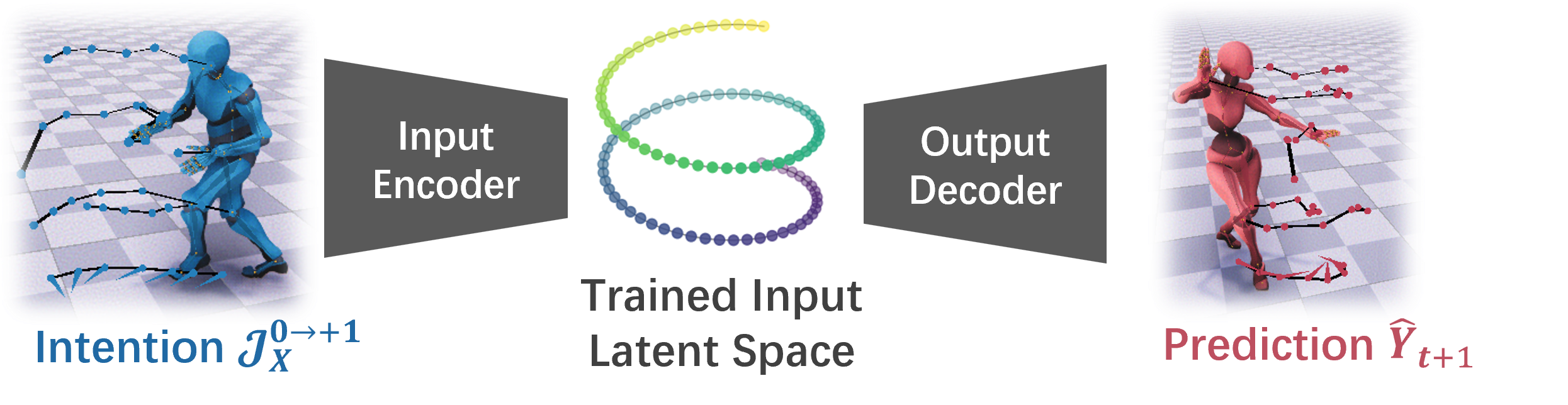}
  \caption{During inference, the predicted intention will be encoded and mapped to a trained latent space from which the Output Decoder predicts the next pose.}
  \label{Inference}
\end{figure}

\section{Literature Reviews}
\subsection{Deep Learning based Motion Synthesis}
Since \cite{holden2015learning,holden2016deep} applied convolutional neural networks (CNNs) to motion data, deep learning-based motion synthesis was widely explored with various network architectures. One main stream of previous studies was to treat motion synthesis as a generative problem and apply generative models to learn the distribution of a whole motion sequence. \cite{guo2024momask} employed a variational autoencoder to model the distribution of human motions. Additionally, normalizing flows were also used to model motion dynamics \cite{henter2020moglow}. Diffusion models \cite{yang2023diffusurvey,chang2025design} were recently widely used for motion synthesis \cite{chang23unifying,cohan2024flexible}. Despite their prosperity, they usually could not be applied to generate long motion sequences. Another method family of motion synthesis focused on auto-regressive generation. By formulating motion synthesis as a sequential generation problem, various recurrent neural networks were proposed to model the temporal dynamics of motions and generate motion sequences frame by frame. \cite{holden2017phase} proposed PFNN for locomotion generation, where the phase representations were used to capture temporal relationships in simple human locomotion. \cite{starke2020local} proposed to model not only the temporal relationships but also the spatial features, and they used the local phase for each joint. \cite{starke2022deepphase} proposed to transform human motions into phase space and apply frequency-based methods to model the periodic dynamics of human motions. However, the previous methods mainly focused on single-character synthesis and did not consider the interactive dynamics within multi-character interactions.

\subsection{Multi-character Motion Synthesis}
Research on multi-character synthesis made significant strides in enabling the generation of realistic interactions between multiple human characters in dynamic environments. Approaches such as \cite{wang2021multi} and \cite{xu2023stochastic} focused on predicting and forecasting complex interactions and body movements among multiple agents. These models handled the intricacies of spatial and temporal relationships between multiple characters. Transformers, in particular, played a pivotal role in capturing multi-person interactions through hierarchical and trajectory-aware frameworks, as seen in \cite{peng2023trajectory}. Additionally, GAN-based reactive motion synthesis \cite{men2022gan} applied class-aware discriminators to enhance the quality of synthesized motion between characters, improving the realism of human-human interaction. Emerging techniques such as diffusion models \cite{tanke2023social,liang2024intergen,chang2025large} further explored long-term multi-human motion anticipation and diffusion-based multi-agent generation, respectively, showcasing the increasing capacity to model complex and natural motion sequences.

While existing approaches primarily focused on generating two-character motions, some studies shifted their attention to predicting the reactive motions of a single character, where interaction modeling was also essential. \cite{chopin2023interaction} was a pioneering work in this area, employing an attention-based network to model both temporal and spatial dependencies. \cite{ghosh2024remos} utilized diffusion models and introduced a reaction loss to enhance prediction accuracy. Similarly, \cite{xu2024regennet} adopted diffusion as the generative model and proposed a distance-based interaction loss. Unlike these methods, which generated uncontrollable reactive motions in an offline manner by leveraging future information of the input character, our work addresses the more challenging online task, requiring accurate and controllable reaction predictions in real time.

\subsection{Real-time Character Control}
While matching-based real-time prediction advanced significantly in recent years—such as in virtual reality (VR) applications—their task was fundamentally different from ours in terms of responsive motion modeling and access to future information. Among these efforts, \cite{lee2023questenvsim} introduced a physics-based approach for motion tracking using sparse inputs from VR devices, enabling real-time motion synthesis with environmental interactions. Similarly, \cite{starke2024categorical} addressed motion prediction from sparse VR inputs by estimating future joint trajectories and learning mappings between these trajectories and output poses through a codebook matching technique. \cite{ponton2022combining} leveraged motion matching combined with predicted root orientation to animate avatars in VR environments. 
While these works shared a similar challenge with our task—namely, addressing ambiguity in prediction—they were fundamentally different. Their task did not require consideration of dynamic interactions between two characters. Specifically, their predictions were conditioned on sparse VR device inputs from a single user, whereas our predictions are constrained by the historical motion of an input character with an unknown future trajectory. This introduces a greater challenge in motion prediction, as achieving realistic interactions requires effective modeling of the relationship between two characters.

\cite{cenready} proposed an auto-regressive online reaction policy that enabled two characters to interact in real time by independently generating motion based on both their own and their counterpart’s motion history. However, their method was not evaluated on broader motion types beyond boxing and lacked flexibility in handling user-controlled input actions. In contrast, we focus on real-time single-character reactive motion synthesis that responds directly to controllable inputs (i.e., user-specified trajectories).

\begin{figure}[tbp]
  \centering
  \includegraphics[width=\linewidth]{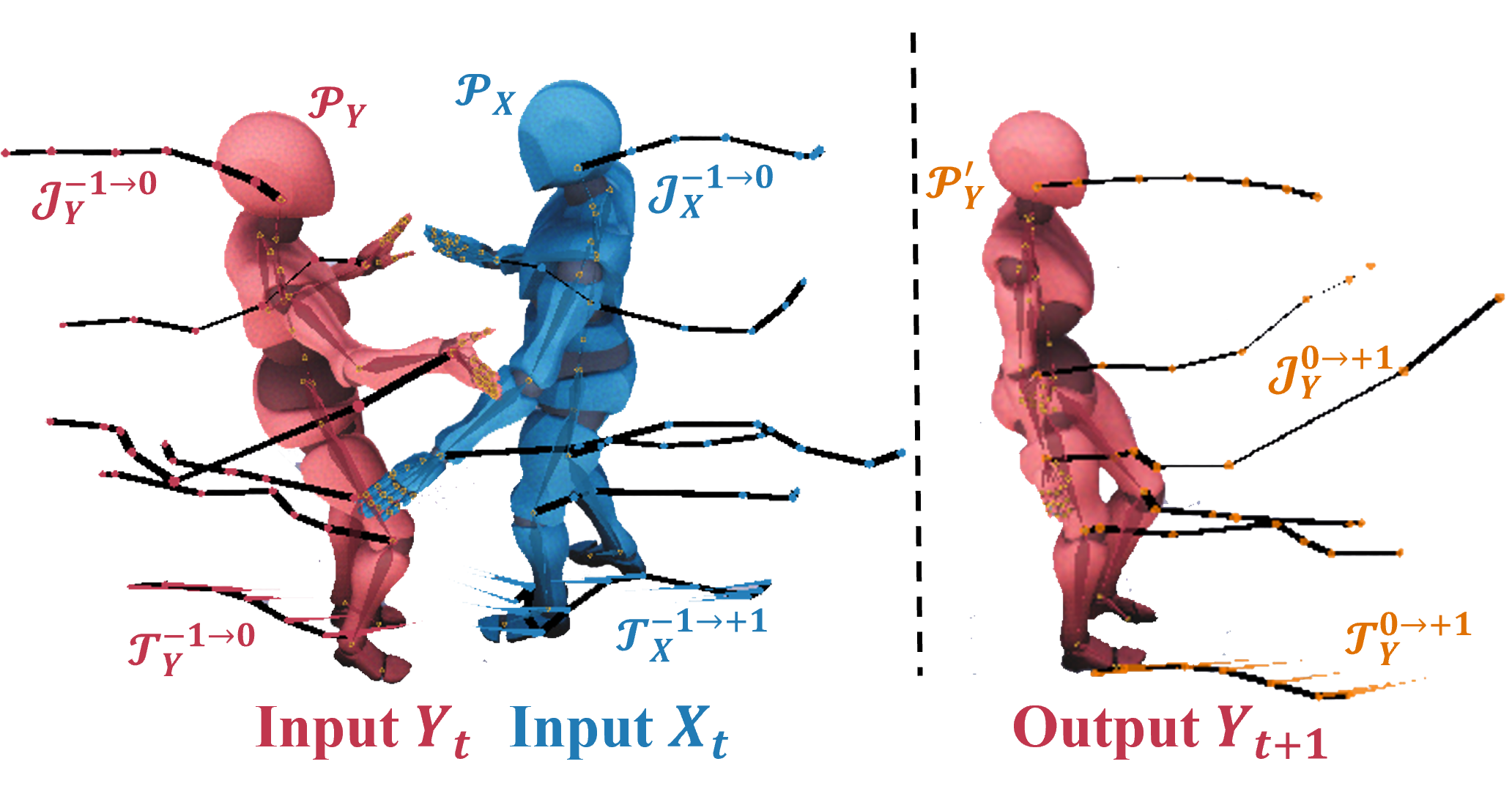}
  \caption{Data representation of input and output. The black lines behind characters represent their historical joints and root trajectories, while the lines in front of the character are the future trajectories.}
  \label{DrawRepresentation}
\end{figure}

\section{Methodology}
Our objective is to predict the reactive pose of character Y at frame $t+1$ based on the observed poses of characters X and Y at frame $t$, formalized as: $Y_{t+1} = f(X_t, Y_t)$.
To generate reactive motions in response to dynamic inputs, our model incorporates two key components.
First, we introduce an intention predictor that estimates future dynamics to stabilize the prediction process.
Second, we propose an adversarial codebook matching mechanism to effectively predict the next pose from the estimated future state.
An overview of the training and inference pipelines is illustrated in Figures~\ref{Architecture} and~\ref{Inference}.

\subsection{Representation} \label{Representation}
We represent the conditioned input character as $X_t$ and the predicted character as $Y_{t+1}$ (see Figure \ref{DrawRepresentation}). Following \cite{starke2020local}, we convert all pose information into local space. The trajectories are represented as $k=6$ keyframes within the past 1 second of time window. The input data includes the following features:
\begin{itemize}
    \item Root trajectories of both characters: $\mathcal{T}^{-1 \rightarrow +1}_{X_t}$ and $\mathcal{T}^{-1 \rightarrow 0}_{Y_t}$, where $\mathcal{T} \in \mathbb{R}^{4 \times k}$ represents the horizontal 2D position and forward direction for each second. The input character $X$ includes an additional 1-second future trajectory to enable user control and directional steering.
    \item Joint trajectories of five key joints (Head, Left Hand, Right Hand, Left Leg, Right Leg) for both characters: $\mathcal{J}^{-1 \rightarrow 0}_{X_t}$ and $\mathcal{J}^{-1 \rightarrow 0}_{Y_t}$, where $\mathcal{J} \in \mathbb{R}^{3 \times k}$ contains the 3D positions of joints for each second.
    \item Pose in current root space: $\mathcal{P}_{X_t}$, $\mathcal{P}_{Y_t}$, where $\mathcal{P} \in \mathbb{R}^{n \times 12}$; $n$ is the number of joints. Each joint includes 3D position, 3D velocity (for smooth translation), and forward/upward vectors representing joint rotation \cite{zhang2018mode}.
\end{itemize}
The output data includes the following features:
\begin{itemize}
    \item Root update: $\mathcal{R}_{Y_{t+1}} \in \mathbb{R}^{3}$, representing the positional and angular offset on the horizontal plane from the previous frame.
    \item Root trajectory over a 1-second future window: $\mathcal{T}^{0 \rightarrow +1}_{Y_{t+1}}$.
    \item Joint trajectory over a 1-second future window: $\mathcal{J}^{0 \rightarrow +1}_{Y_{t+1}}$.
    \item Pose in the updated root space: $\mathcal{P}_{Y_{t+1}}$.
    \item Foot contact features for the next frame: $\mathcal{C}_{Y_{t+1}} \in \mathbb{R}^{2}$, used to reduce foot sliding artifacts via inverse kinematics \cite{starke2020local}.
\end{itemize}
For simplicity, we denote all input at frame $t$ and output at frame $t+1$ as:
\begin{equation}
    X_t=\{ \mathcal{T}^{-1 \rightarrow +1}_{X_t}, \mathcal{J}^{-1 \rightarrow 0}_{X_t}, \mathcal{P}_{X_t}\},
\end{equation}
\begin{equation}
    Y_t=\{ {\mathcal{T}}^{-1 \rightarrow 0}_{Y_t}, {\mathcal{J}}^{-1 \rightarrow 0}_{Y_t}, \mathcal{P}_{Y_t}\},
\end{equation}
\begin{equation}
    Y_{t+1} = \{ \mathcal{R}_{Y_{t+1}}, \mathcal{T}^{0 \rightarrow +1}_{Y_{t+1}}, \mathcal{J}^{0 \rightarrow +1}_{Y_{t+1}}, \mathcal{P}_{Y_{t+1}}, \mathcal{C}_{Y_{t+1}} \}.
\end{equation}
Inputs and outputs are formed as $\mathcal{I} = [X_t; Y_t] \in \mathbb{R}^{1504}$ and $\mathcal{O} = [Y_{t+1}] \in \mathbb{R}^{743}$

\subsection{Intention Predictor}
Predicting reactive motions in real-time can be highly noisy and ambiguous, as the reactor lacks knowledge of the input character's future movements. This uncertainty can degrade the quality of interaction in real-time synthesis, as errors may accumulate over time.

To enable responsive reaction synthesis, we predict the next reactive pose from the estimated \textit{intention} of the input actor where the intention is represented as the future joint trajectories transformed into the local space of the reactor's root. 
This intention feature is crucial as it provides prior knowledge of future interactions, enabling the model to learn an effective mapping between the output pose and the predicted future trajectories rather than relying solely on observed history. Without this feature, the same inputs could be encoded as ambiguous codebook vectors, leading to noisy and inconsistent poses.

Given the input features in each frame as stated in Section \ref{Representation}, we pre-train an intention predictor consisting of fully-connected layers to predict the future trajectories of the input character. Instead of anticipating all joints, we only predict the intention movement on the selected five key joints, as described in Section \ref{Representation}. The dynamics of the five joints are representative of the overall body movement while maintaining a manageable feature dimensionality for efficient neural network training: 
\begin{equation}
    IntentionPredictor(X_t, Y_t) = \hat{\mathcal{J}}^{0 \rightarrow +1}_{X_t},
\end{equation}
where $\hat{\mathcal{J}}^{0 \rightarrow +1}_{X_t}$ is the intention (i.e., the future joint trajectories) of character $X$.

The intention predictor is trained by minimizing the Mean Squared Error (MSE) between predicted intention $\hat{\mathcal{J}}^{0 \rightarrow +1}_{X_t}$ and ground truth $\mathcal{J}^{0 \rightarrow +1}_{X_t}$:
\begin{equation}
    \mathcal{L}_{\text{Intention}} = MSE(\mathcal{J}^{0 \rightarrow +1}_{X_t}, \hat{\mathcal{J}}^{0 \rightarrow +1}_{X_t}).
\end{equation}

\subsection{Reactive Motion Synthesis}
\subsubsection{Adversarial Codebook Matching}
Our approach builds upon Codebook Matching \cite{starke2024categorical} for motion synthesis. This method learns to align the latent distributions of per-frame input and output features, enabling the synthesis process to be interpreted as reconstructing outputs through an autoencoder network. To extend codebook matching to reactive motion between two characters, we adapt the original framework of \cite{starke2024categorical}, which was designed for single-character prediction.

Unlike their setup that maps future trajectories to poses based solely on self-motion, our task requires modeling the interactive dynamics between two characters. A key challenge in this setting is that interaction depends on the \textit{relative} motion between characters. To capture this, we transform the future trajectories of the input character into the local coordinate space of the reactor’s root (see Figure~\ref{Representation}). This allows the model to focus on relational dynamics rather than absolute movement.

However, the increased diversity introduced by relative representations poses a significant challenge: the same interaction may appear differently depending on the reactor’s root orientation and motion. This variability makes it difficult for a discrete codebook to adequately cover the continuous space of interactions, often leading to mismatches and ambiguous predictions.

To address this, we propose to complement the codebook matching objective with adversarial training in the continuous latent space. Specifically, we align the latent encoding of the input’s estimated intention with the latent representation of the output reactions using adversarial loss. This dual-objective training promotes both (1) accurate quantization into the codebook vector, and (2) a well-structured latent manifold that preserves relational semantics and improves generalization.

Without explicit alignment in the continuous latent space, the model may incorrectly assign distinct interaction patterns to the same codebook vector or distribute similar interactions across different vectors, resulting in unstable or incoherent motions. In contrast, our adversarially-regularized latent space offers greater robustness in encoding diverse two-character interactions, enhancing the model’s ability to synthesize accurate and coherent reactive motions across a variety of scenarios.

\begin{figure}[tbp]
  \centering
  \includegraphics[width=\linewidth]{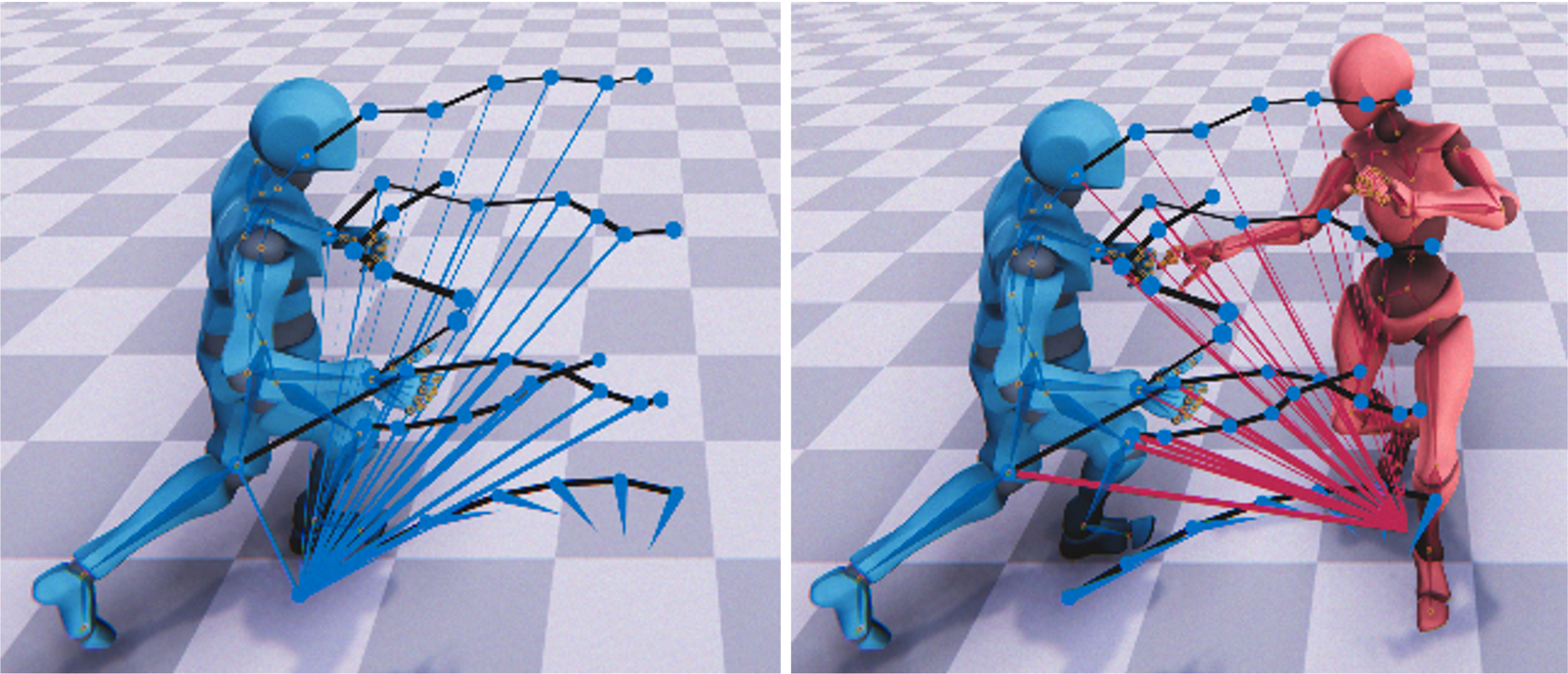}
  \caption{Visual comparison of different trajectory representations. The left example shows future trajectories represented in the character’s own root space, while the right example uses the reactor’s root space as the reference frame. Accurate interaction prediction in our task relies on observing relative spatial information, presenting greater challenges than those encountered in single-character motion prediction.}
  \label{Representation}
\end{figure}

Specifically, we encode the input and output features $\mathcal{I}$, $\mathcal{O}$ and extract the latent distribution, where we aim to align the estimated distribution $\hat{\mathcal{S}}$ with the target one $\mathcal{S}$:
\begin{equation}
    InputEncoder(\mathcal{I}, \hat{\mathcal{J}}^{0 \rightarrow +1}_{X_t}) = \hat{\mathcal{S}},
\end{equation}
\begin{equation}
    OutputEncoder(\mathcal{O}, \mathcal{I}, \hat{\mathcal{J}}^{0 \rightarrow +1}_{X_t}) = \mathcal{S}.
\end{equation}
A categorical probability distribution is sampled from the latent space of input encoder and output encoder through Gumbel-Softmax method (denoted as $GS$)\cite{jang2017categorical} where each category represents a pose update responsive to the input interactions. The codebook vectors $\mathcal{B}\in \mathbb{R}^{c\times d}$ with $c$ channels and $d$ dimensions are quantized using differentiable argmax and straight-through estimator (STE):
\begin{equation}
    \mathcal{B} = GS(\mathcal{S}),
\end{equation}
\begin{equation}
    OutputDecoder(\mathcal{B}) = \hat{\mathcal{O}}.
\end{equation}

\subsubsection{Training}
The original codebook matching loss enforces consistency at the quantized level by aligning discrete codebook vectors between input and output encoders. In contrast, adversarial training encourages alignment in the continuous latent space before quantization. These two mechanisms are complementary: one ensures accurate codebook embedding, while the other promotes a well-structured latent manifold that improves matching stability and diversity.

The output auto-encoder is trained by minimizing the MSE between reconstruction $\hat{\mathcal{O}}$ and ground truth output $\mathcal{O}$:
\begin{equation}
    \mathcal{L}_{\text{Reconstruction}} = MSE(\mathcal{O}, \hat{\mathcal{O}}).
\end{equation}
The codebook is optimized through minimizing the MSE between target codebook $\mathcal{B}$ and estimated codebook $\hat{\mathcal{B}}$:
\begin{equation}
    \mathcal{L}_{\text{Codebook}} = MSE(\mathcal{B}, \hat{\mathcal{B}}).
\end{equation}
The adversarial loss is formulated as:
\begin{equation}
    \mathcal{L}_{\text{D}} = - \mathbb{E}_{\mathcal{S} \sim p_(\mathcal{S})} [\log D(\mathcal{S})] - \mathbb{E}_{\hat{\mathcal{S}} \sim p_(\hat{\mathcal{S}})} [\log (1 - D(\hat{\mathcal{S}}))],
\end{equation}
\begin{equation}
    \mathcal{L}_{\text{G}} = - \mathbb{E}_{\hat{\mathcal{S}} \sim p_(\hat{\mathcal{S}})} [\log D(\hat{\mathcal{S}})],
\end{equation}
\begin{equation}
    \mathcal{L}_{\text{GAN}} = \mathcal{L}_{\text{D}} + \mathcal{L}_{\text{G}},
\end{equation}
where $D$ is the discriminator, $\mathcal{L}_{\text{D}}$ and $\mathcal{L}_{\text{G}}$ are discriminator loss and generator loss. Finally, the loss function is formed as $\mathcal{L} = \lambda_R\mathcal{L}_{\text{Reconstruction}} + \lambda_C\mathcal{L}_{\text{Codebook}} + \lambda_G\mathcal{L}_{\text{G}}$ with weights applied respectively.
We empirically investigate the effect of the applied weight balance in Section \ref{Effectiveness of Adversarial Codebook Matching}, demonstrating that the combination enhances the robustness of codebook matching.


\section{Experiments}
\subsection{Implementation Details}
Our neural network is trained on a public ReMoCap dataset \cite{ghosh2024remos}, a synthetic boxing dataset \cite{shum2010simulating, shum07simulating} and the DD100 dataset \cite{siyao2024duolando}. They contain 203502, 50635 and 42832 frames of augmented motion capture data in 30 FPS. The proportions of the training set, validation set, and test set are 90\%, 5\%, and 5\%, respectively. We operate motion re-targeting to fit in the skeleton of Mixamo \cite{mixamo} character with 52 joints. 

The visualization system is adapted from the open-source Unity3D motion animation system \cite{starke2020local} and uses socket communication to transfer motion data between Unity3D engine and the neural network. The inference time is approximately 7 ms per frame at 30Hz.

The intention predictor and adversarial codebook matching module each consist of three fully connected layers, while the discriminator comprises two fully connected layers. The dimension of hidden layer is set to be 512 and 2048 for intention predictor and adversarial codebook matching, respectively. We choose channel $c=16$ and dimension $d=128$ as the size of codebook. 
The number of codebook channels is increased compared to the original setting in \cite{starke2024categorical}, as our model encodes the concatenated motion features of two interacting characters. Each character has 52 joints, compared to 27 joints in \cite{starke2024categorical}. This expansion is necessary to accommodate the greater complexity and dimensionality of the input representation. 
To ensure a fair comparison, we also evaluate the original codebook matching method with an increased parameter size (denoted as \cite{starke2024categorical}(L)). The architectural details and parameter configurations are provided in Figure~\ref{EncoderDetail} and Table~\ref{Parameters}.

The entire model, including the intention predictor, is trained for 150 epochs with a batch size of 16. Training takes approximately 15 to 20 hours depending on the dataset size, conducted on NVIDIA RTX 3080 GPU. 
Following \cite{starke2024categorical}, we use the AdamW optimizer \cite{loshchilov2018decoupled} with an initial learning rate of $10^{-4}$ and apply a cosine annealing schedule for learning rate decay.

\begin{figure}[tbp]
  \centering
  \includegraphics[width=\linewidth]{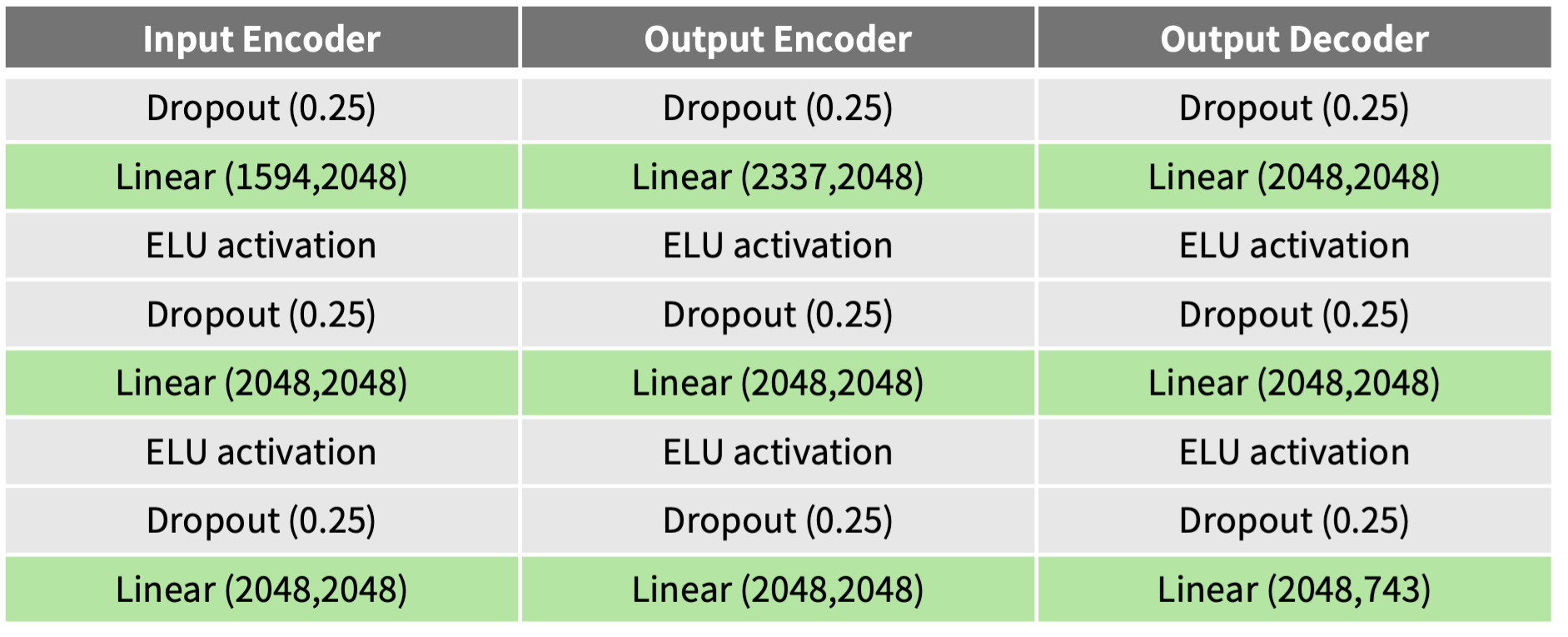}
  \caption{Architecture details of the adversarial codebook matching.}
  \label{Encoder Detail}
\end{figure}
\begin{table}[!ht]
    \centering
    \begin{tabular}{l|c|c|c}
        ~ & Ours & \cite{starke2024categorical} & \cite{starke2024categorical}(L) \\
        \hline
        Channels & 16 & 8 & 24 \\
        Dimension of channel & 128 & 128 &  128 \\ 
        Codebook size & 2048 & 1024 & 3072 \\ 
        Total parameters & 36 million & 12 million & 72 million \\ 
        Memory & 136 MB & 46 MB & 270 MB \\ 
    \end{tabular}
    \caption{Comparison of parameter configurations across different models. Total parameters in our model includes the pre-trained intention predictor. \cite{starke2024categorical}(L) denotes the original codebook matching model with a larger capacity.}
    \label{Parameters}
\end{table}

\begin{figure*}[tbp]
  \centering
  \includegraphics[width=\linewidth]{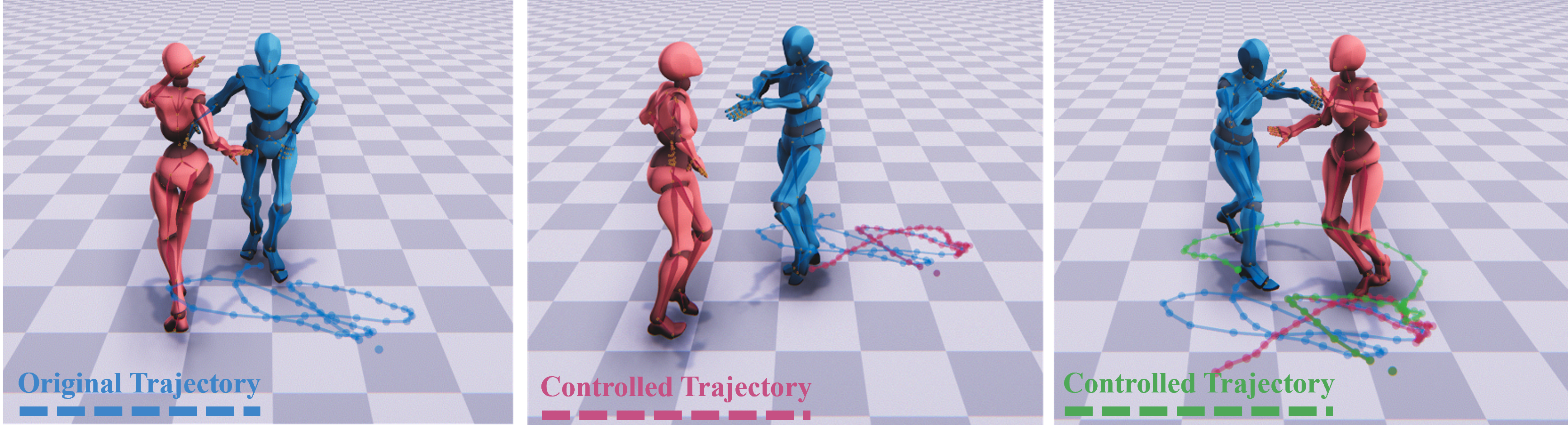}
  \caption{Left: Performance using original motion from the testing dataset as input. Middle and Right: Performance using controlled motion (red and green trajectories) as input.}
  \label{DrawHistory}
\end{figure*}

\begin{figure}[tbp]
  \centering
  \includegraphics[width=\linewidth]{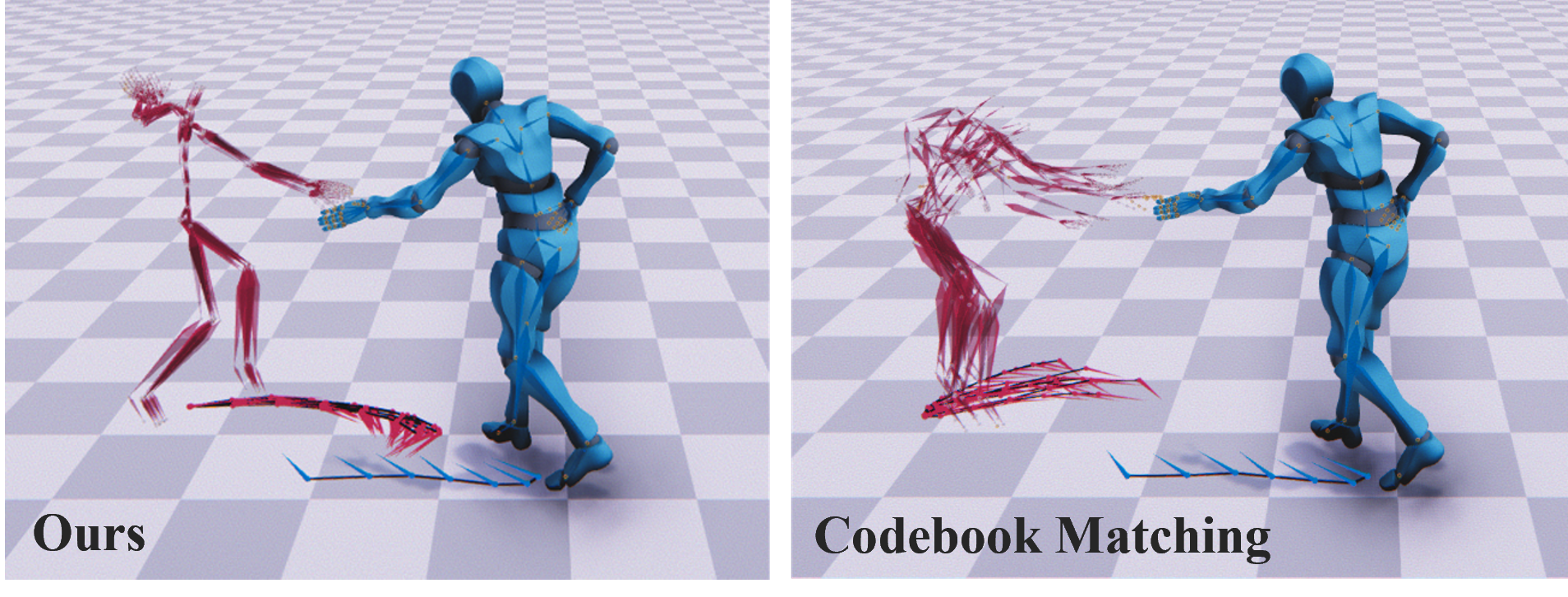}
  \caption{For better visualization, we remove character's mesh to showcase the variation of different samples. Left: Our model produces stable predictions across all samples. Right: \cite{starke2024categorical} tends to generate noisy results due to the high uncertainty in predicting the input character's future trajectory.}
  \label{DrawSamples}
\end{figure}

\subsection{Evaluation Metrics and Comparison Methods}
We first use Mean Per Joint Position Error (MPJPE) to assess positional errors relative to the ground truth. However, since MPJPE alone is insufficient for evaluating the quality of the generated motions, we also utilize Frechet Inception Distance (FID) \cite{heusel2017gans} and the latent variance of generated motions (Diversity) \cite{shafir2024human} to analyze the generation quality through high-dimensional latent features in the latent space. For both Diversity and FID, we extract latent features directly from the encoder output. Additionally, to assess the network's ability to map the input to a valid output among multiple ambiguous options, we calculate the Mean Squared Error (MSE) between the target codebook from the Output Encoder and the estimated codebook from the Input Encoder, representing the codebook matching error.

For a widely-used standard comparison, we test the performance of basic Motion Matching approach \cite{gdcvault2024motionmatching,holden2020learned} as an evaluation benchmark, where we directly search for the next pose within the training dataset, using the root trajectory $\mathcal{T}^{0 \rightarrow +1}_{X_t}$, $\mathcal{T}^{0 \rightarrow +1}_{Y_t}$, as well as the poses $\mathcal{P}_{X_t}$ and $\mathcal{P}_{Y_t}$, as query features. We also test the performance by applying codebook matching \cite{starke2024categorical} on two-character interaction synthesis, encoded solely on historical interaction. The performance is tested under the same latent space dimensions for fair comparison, as well as under a larger model where dimension tripled. 

\subsection{Applications}
We conduct a throughout experimental test to evaluate our reactive motion generation method from the following aspects: 
\textbf{Real time, Accuracy} - We first show the feasibility of our method in generating long-term, high-quality dancing interaction;
\textbf{Controllable} - We further show how the method adapt to personalized dancing path through interactive user control;
\textbf{Guidance} - In ablation study, we show how the intention guidance helps stabilize the reactive motion space. We also evaluate how adversarial training on codebook matching can positively improve the learned motion latent space.


\begin{table}[!ht]
    \centering
    \begin{tabular}{l|c|c|c|c}
        ~ & MPJPE$\downarrow$ & Diversiy$\rightarrow$ & FID$\downarrow$ & Codebook$\downarrow$ \\
        \hline
        Ground Truth & - & 85.35 & - & - \\
        MM & 1.88 & 81.65 & 0.47 & 0.011 \\ 
        \cite{starke2024categorical} & 7.74 & 57.78 & 46.60 & 0.057 \\ 
        \cite{starke2024categorical}(L) & 7.69 & 58.34 & 44.90 & 0.055 \\ 
        Ours & 6.91 & 61.77 & 28.36 & 0.048 \\ 
    \end{tabular}
    \caption{Quantitative results without control signals applied. $\downarrow$: Lower is better. $\rightarrow$: Closer to ground truth is better. (L): With larger dimension of latent features.}
    \label{InfiniteSynthesisQuantitative}
\end{table}

\subsubsection{Infinite Reactive Motion Synthesis}
Our method effectively generates long-term, vivid, and responsive motion by alternatively estimating the future movements of the two characters. 
We demonstrate the diversity of prediction by sampling multiple times and visualize the results in Figure \ref{DrawSamples}. Our results show stable and reasonable sampling performance, with an average inference time of 7 ms per frame.

Meanwhile, the output of \cite{starke2024categorical} displays significantly higher levels of noise (as shown in Figure \ref{DrawSamples}, right). This is largely due to that the model, when given only historical motion input, maps it to multiple potential codebooks, resulting in greater uncertainty and instability in the generated output. Unlike the relatively simple single-person motions typically used in \cite{starke2024categorical}, handling two-person dancing motion involves more dynamic interactions constraints. 
The lack of future pose information in \cite{starke2024categorical} increases uncertainty in predicting actions, raising the risk of incorrect or unrealistic interactions.

From a quantitative perspective (see Table \ref{InfiniteSynthesisQuantitative}), we consistently outperform \cite{starke2024categorical} across all metrics, regardless of the parameter size. Inferior codebook matching error highlights the difficulty of capturing the full range of possible mappings between input and output poses. A larger model capacity (\cite{starke2024categorical}(L)) may help mitigate this, but it is still limited by the invisibility to the potential future knowledge, making it much worse than predicting an intention prior in advance as in our method. Practically, a large-scale and complex synthesis model will also slow down the real time performance. Additionally, we maintain a higher Diversity score with a lower FID score, indicating the generation of more realistic and varied motion. 

On the other hand, we observe that motion matching performs well by identifying closely aligned poses from the training dataset with minimal deviation from the ground truth (see MM in Table \ref{InfiniteSynthesisQuantitative}). The method’s ability to deliver such accurate results stems from its efficiency in locating similar poses within the dataset, keeping errors relatively small. However, the main limitation of employing MM in our real-time dancing motion synthesis lays in two fields: (1) Selected from the pre-defined dataset of animations, the retrieved dancing motion lack of variations and not generalizable to any new interaction patterns from user input, which we will demonstrate in Section \ref{ControllableReactiveMotionSynthesis}. 
(2) It is memory-intensive to store large amount of animation and computational expensive to process the data efficiently, which makes it not practical for real-time dynamics generation.

\begin{figure}[tbp]
  \centering
  \includegraphics[width=\linewidth]{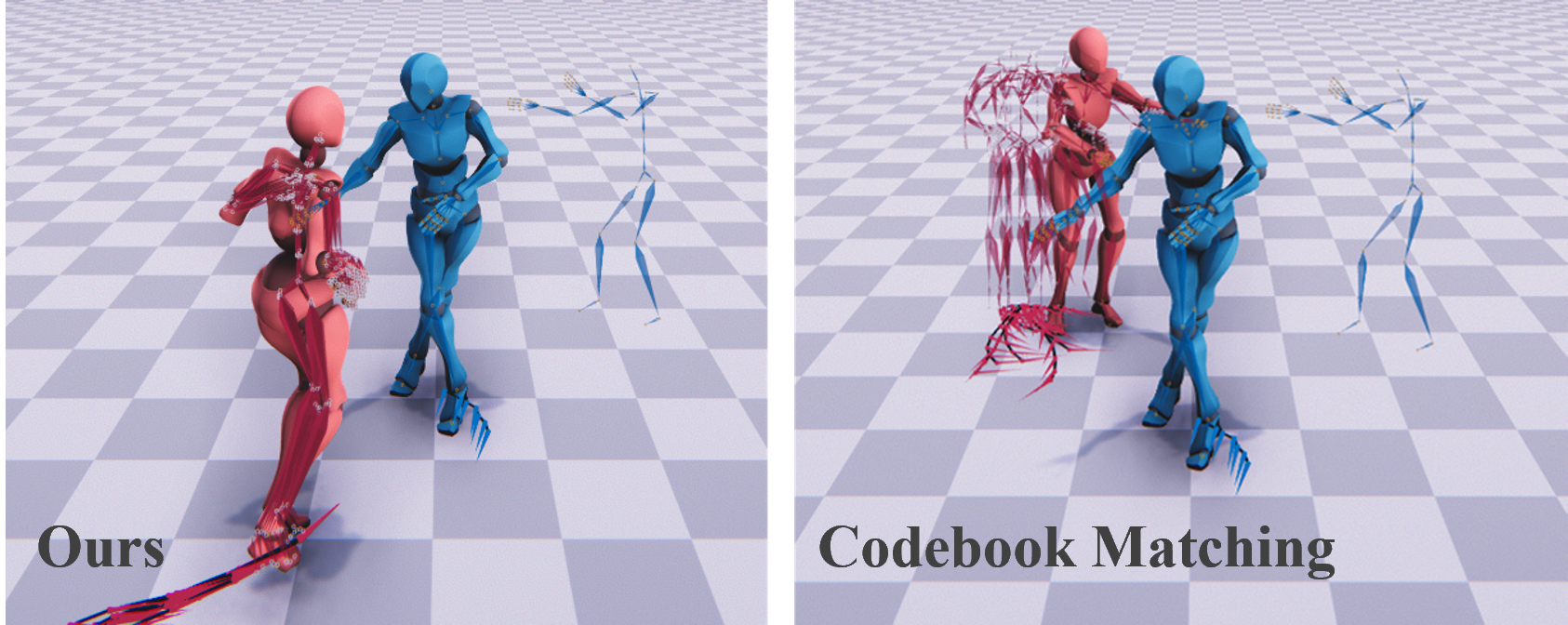}
  \caption{The blue skeleton without the mesh represents the original ground truth position, while the blue character with the mesh shows the position after controls are applied. Left: Our model generates valid and interactive motions under controlled conditions. Right: \cite{starke2024categorical} quickly collapse, producing noisy and non-interactive motions.}
  \label{DrawControl}
\end{figure}


\subsubsection{Controllable Reactive Motion Synthesis}\label{ControllableReactiveMotionSynthesis}
We show the generalizability of our system adapted to real-time user control of the interaction direction. This is especially useful for creating dynamic dance movements by controlling directions like forward, backward, left, right, and diagonal, resulting in more active and vivid performances.
Specifically, we allow users to modify the future root trajectory (blue character in Figure 5), $\mathcal{T}^{0 \rightarrow +1}_{X_t}$, using keyboard inputs as control signals and search for the next pose in the database which matches the controlled root trajectories. The character orientations and local limb movement will change to the next pose accordingly. This allows the motion to deviate dynamically and naturally from its original path and form an interaction condition that does not exist in the dataset.

Figure \ref{DrawHistory} illustrates the results of this user-controlled experiment, showing the differences between the original trajectory (left subfigure) and two example modified trajectories (subfigures in middle and right) under user control. In each case, the poses of character $X$ adapt to the new control inputs, successfully generating realistic and coherent motion that reflects the changes in direction induced by the user. Even when real-time input states, such as the relative positions of characters or their historical trajectories, deviate significantly from the ground truth, the network continues to produce visually plausible motions. This highlights the model’s robustness and its ability to handle variations that were not explicitly encountered during training. 

In contrast, baseline methods (\cite{gdcvault2024motionmatching} and \cite{starke2024categorical}) struggle when user-controlled signals are introduced. These models tend to produce erratic or frozen motions that fail to interact properly with the modified trajectories. Figure \ref{DrawControl} (right) and the supplemental video demonstrate the breakdown of these methods under user control, resulting in noisy or static outputs that lack fluidity and responsiveness. We argue that Motion Matching is heavily reliant on matching pre-defined poses from a limited database of possible motions, and it lacks the flexibility to adapt dynamically to control inputs that push the model outside its pre-existing motion space. While a codebook may be sufficient to capture all possible pose combinations in a single-character scenario, it struggles to adapt in two-character scene, where the interaction patterns such as the relative positioning between characters are far more dynamic and different from preset motion sequences in dataset. 

\begin{table}[!ht]
    \centering
    \begin{tabular}{l|c|c|c}
        ~ & MPJPE$\downarrow$ & FID$\downarrow$ & Codebook$\downarrow$ \\
        \hline
        Complete Setup & \textbf{6.91} & \textbf{28.36} & \textbf{0.048} \\ 
        w/o intention predictor & 7.79 & 43.77 & 0.063 \\ 
        End-to-end & 7.04 & 30.89 & 0.050 \\ 
        End-to-end (w/o SS) & 7.70 & 40.96 & 0.065 \\
        w/o GAN & 7.13 & 34.03 & 0.054 \\ 
    \end{tabular}
    \caption{Quantitative results compared with ablated versions or end-to-end version of intention predictor.}
    \label{AblationQuantitative}
\end{table}

\subsection{Ablation Study}
\subsubsection{Effectiveness of Intention Predictor}
\begin{figure}[tbp]
  \centering
  \includegraphics[width=\linewidth]{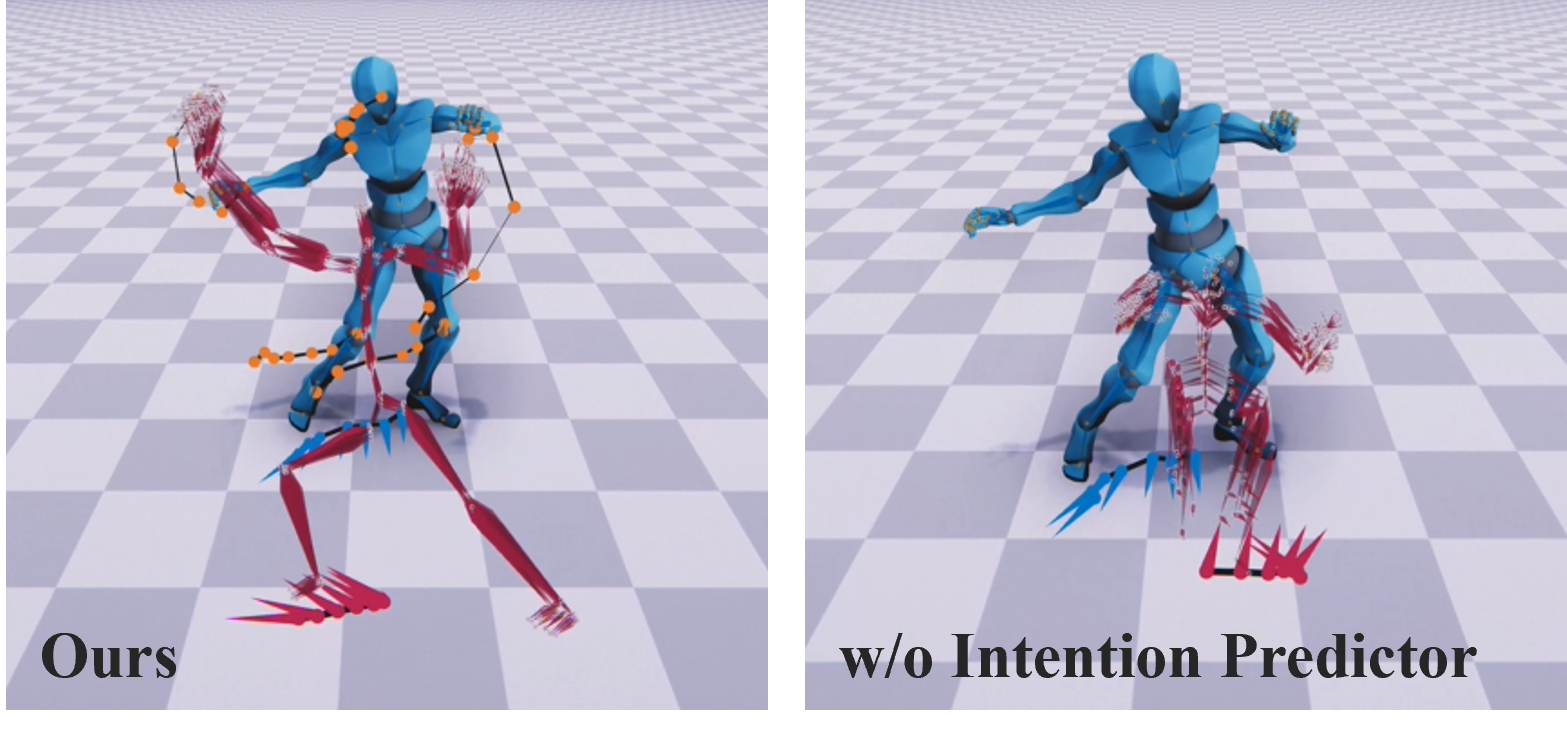}
  \caption{The black lines with orange dots are the predicted intentions. Left: Performance with intention predictor. Right: Performance without intention predictor.}
  \label{AblatedIntention}
\end{figure}

The intention predictor plays a crucial role in guiding and stabilizing reaction synthesis by providing more deterministic conditions for encoding and predicting the next pose. To evaluate its effectiveness, we sample multiple instances per frame and visualize the anticipated intentions and poses. As shown in Figure \ref{AblatedIntention}, our model consistently generates clear and precise future trajectories (illustrated as yellow lines) for both characters. In contrast, the ablated version without intention predictor produces noticeably noisier results for both characters because the inputs are mapped to ambiguous codebook vectors.

\begin{figure}[tbp]
  \centering
  \includegraphics[width=\linewidth]{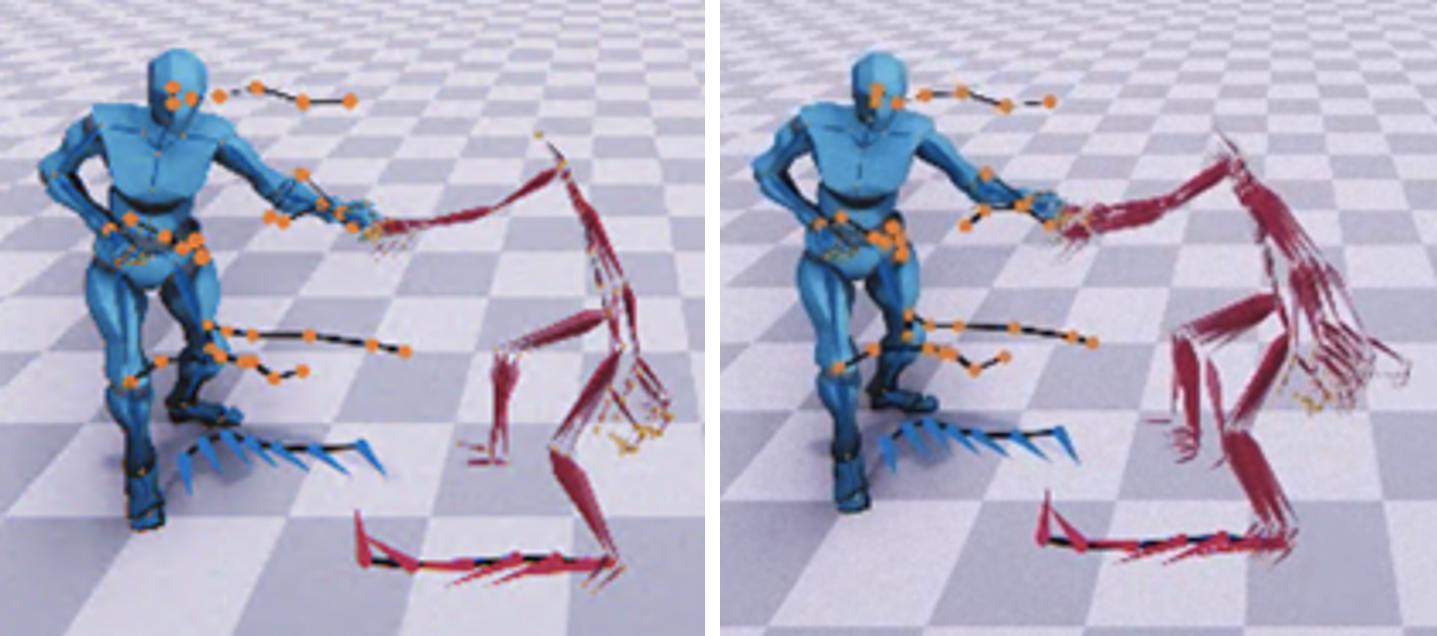}
  \caption{By adjusting the temperature of the Gumbel-Softmax sampling, the model can generate non-deterministic reactive poses (red skeletons) at each step from the same input conditions.}
  \label{PredictionNoise}
\end{figure}

While the intention predictor operates deterministically, it is essential for reducing the ambiguity inherent in future interactions, thereby supporting stable, controllable, and real-time reactive motion synthesis. Nonetheless, the model as a whole remains stochastic, as Gumbel-Softmax sampling introduces controlled randomness into the generation process (See Figure~\ref{PredictionNoise}). In contrast, simply applying original probabilistic codebook matching performs poorly in our task, as it tends to map distinct relative dynamics to the same codebook vector and result in less diverse or unstable predictions (as shown in Table~\ref{InfiniteSynthesisQuantitative} and supplementary video).

We also investigate the impact of the pre-trained intention predictor by training it in an end-to-end manner. As we feed the encoders with the output of intention predictor that has low accuracy at early training stage, we employ scheduled sampling to help the network adapt to its own errors:
\begin{equation}
\begin{aligned}
IntentionPredictor(\mathcal{I}) = \left\{
\begin{array}{ll}
\hat{\mathcal{J}}^{0 \rightarrow +1}_{X_t}, & \text{with probability } p, \\
\mathcal{J}^{0 \rightarrow +1}_{X_t}, & \text{with probability } 1 - p
\end{array}
\right.
\end{aligned}
\end{equation}
where $p = \frac{e}{E}$, $e$ is current training epoch and $E$ is total training epochs. This ensures a fair comparison between the end-to-end version and the pre-trained version. Quantitative results are presented in Table \ref{AblationQuantitative}. Performance without scheduled sampling is also provided (refer to 'End-to-end (w/o SS)').

The pre-trained version demonstrates a significant improvement in motion prediction and codebook alignment compared to the version without the intention predictor. Although scheduled sampling is employed, the end-to-end version still performs slightly worse than the pre-trained model. Qualitative results are available in the supplementary video.


\subsubsection{Effectiveness of Adversarial Codebook Matching}
\label{Effectiveness of Adversarial Codebook Matching}
\begin{figure}[tbp]
  \centering
  \includegraphics[width=\linewidth]{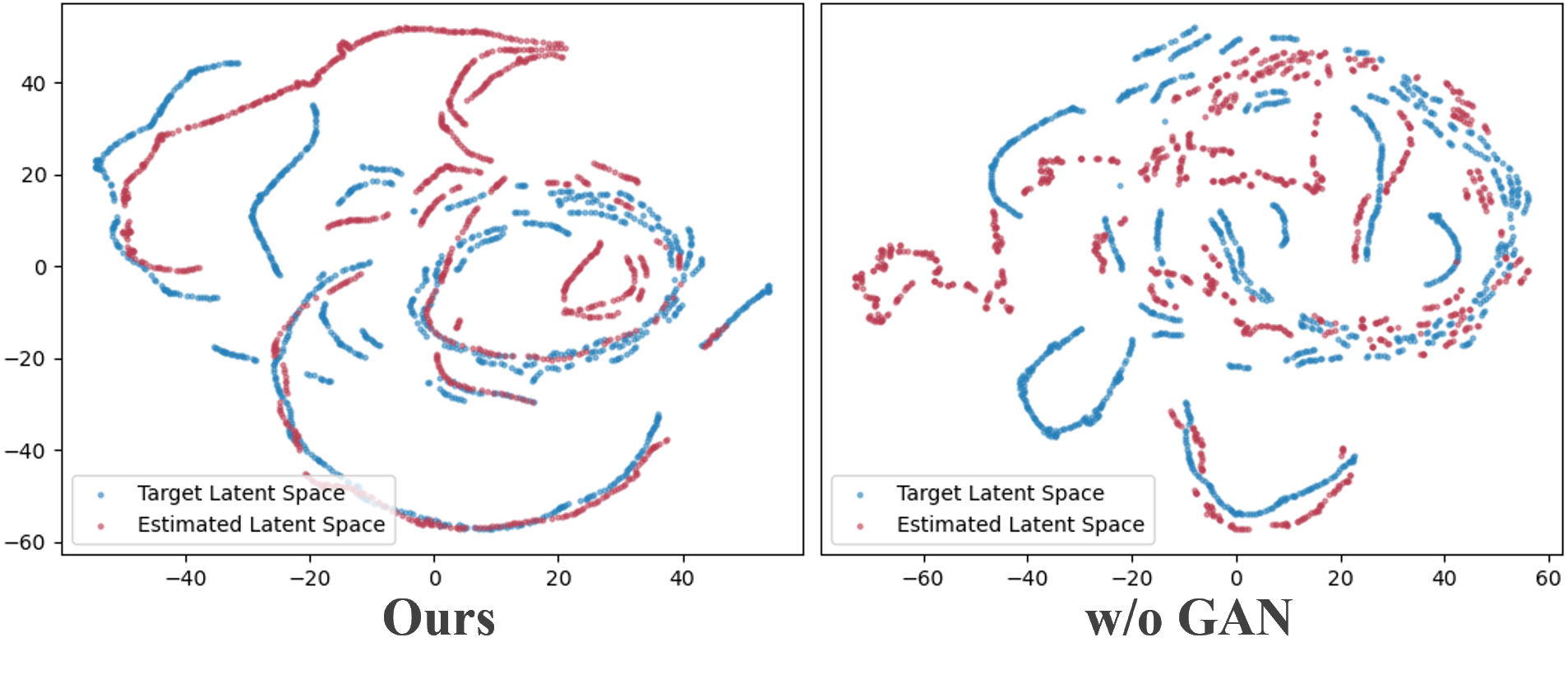}
  \caption{T-SNE results on latent spaces. For better visualization, we only choose 1000 frames for calculation. Blue: Target latent space. Red: Estimated latent space. Left: Result of using adversarial training. Right: Result without adversarial training.}
  \label{DrawLatentSpace}
\end{figure}

\begin{figure}[tbp]
  \centering
  \includegraphics[width=\linewidth]{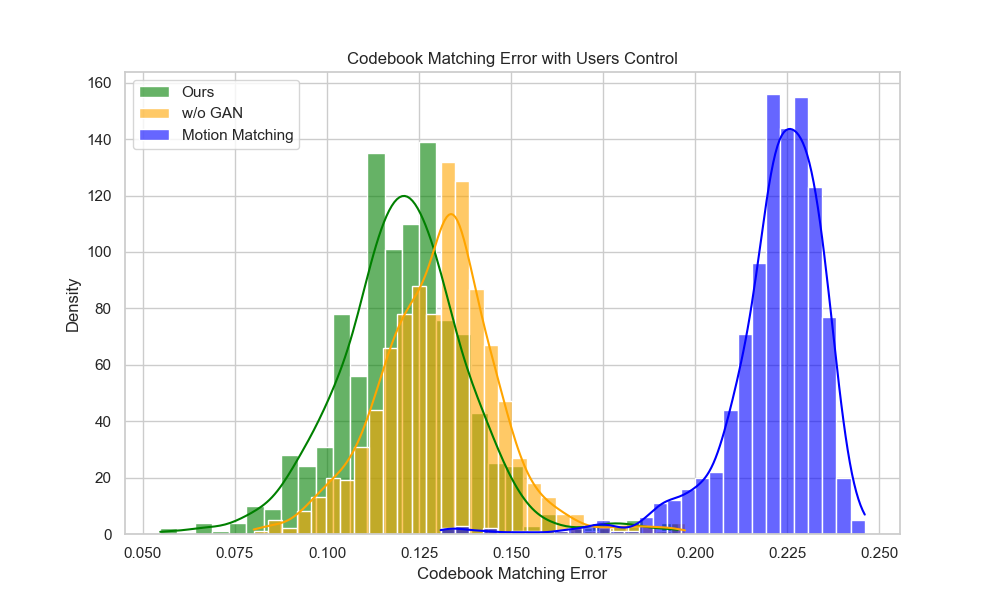}
  \caption{Density of codebook matching error. Adversarial training helps to increase robustness of codebook matching in user-controlled environment while Motion Matching lacks generalizability in this case.}
  \label{CodebookErrorDensity}
\end{figure}

The adversarial training on the motion latent space encourages the network to encode a realistic yet diverse representation of the reactive motion. 
In Table \ref{AblationQuantitative}, our approach achieves an 11\% reduction in codebook matching error with the adversarial training loss (0.048 \textit{vs.} 0.054). 
To illustrate this, we visualized the error between the estimated latent spaces and codebooks against the targets, as shown in Figures \ref{DrawLatentSpace} and \ref{CodebookErrorDensity}. 
Figure \ref{DrawLatentSpace} presents higher overlap and continuity which indicates our GAN enables the encoder to capture a more continuous motion manifold while better aligning with the target latent distribution.
Figure \ref{CodebookErrorDensity} depicts the density of codebook matching errors. Notably, Motion Matching exhibits significantly poorer codebook alignment and generates lower-quality motion. This is primarily due to its inability to dynamically adapt to unseen inputs when user control signals alter the relationships between two characters.

\begin{figure}[tbp]
  \centering
  \includegraphics[width=\linewidth]{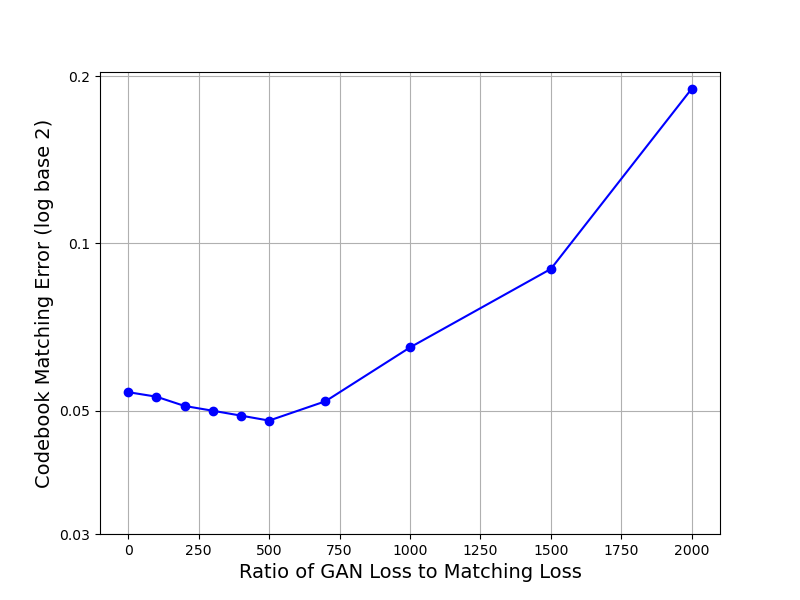}
  \caption{The codebook matching error for various combinations of GAN loss and matching loss.}
  \label{MatchingError}
\end{figure}

We further investigate the tradeoff between adversarial GAN loss and matching loss by adjusting the ratio $\lambda_G / \lambda_C$ between them (Figure \ref{MatchingError}). This exploration is crucial in understanding how the balance between these two losses impacts the overall performance of our model. Notably, we observe optimal performance on codebook matching error when $\lambda_G / \lambda_C = 500$, highlighting the role of GAN in effectively encoding high-dimensional motion data. However, we observe a sharp rise in the matching error as the weight of the adversarial training increases beyond this optimal point. This trend is likely attributable to adversarial training leading to mode collapse, which diminishes the effectiveness of the matching loss, ultimately impairing the training of the encoders.

\section{Limitations}
We have identified several limitations in our method, particularly in how it handles different periods of dancing motion. The predicted motion appears more natural and fluid when the input character is about to perform actions that are inherently periodic, such as two characters facing each other and moving in a circular pattern. However, the method struggles when the input motion is non-periodic or lacks a clear repetitive structure. For example, when a character is standing still and performing actions like clapping, the predicted motion often fails to capture the nuances of these irregular movements, and the output character can appear unnaturally frozen or jittering. This suggests that our method has difficulty adapting to motions with less inherent predictability or regularity, where subtle variations in joints' movement are more critical for producing lifelike motion.

Additionally, since our network predicts the pose and root update separately, foot sliding occurs when users modify the input trajectories, causing the model to generate a responsive root update that is not properly aligned with the full body movement.

\section{Conclusion}
In this paper, we focus on real-time reactive motion synthesis driven by user-defined control signals. We propose an intention predictor to mitigate uncertainty and ambiguity in predicting reactions, and introduce adversarial training into codebook matching to improve latent space alignment. Our model demonstrates strong generalizability, performing well on both the test dataset and customized input trajectories.

\section*{Acknowledgment}
This research is supported in part by the EPSRC NortHFutures project (ref: EP/X031012/1).

\bibliographystyle{eg-alpha} 
\bibliography{Main}

\newcommand{\etalchar}[1]{$^{#1}$}
\begin{thebibliography}{\uppercase{WXNW21}}

\bibitem[CFZS23]{chang23unifying}
\textsc{Chang Z., Findlay E. J.~C., Zhang H., Shum H. P.~H.}:
\newblock Unifying human motion synthesis and style transfer with denoising diffusion probabilistic models.
\newblock In \emph{Proceedings of the 2023 International Conference on Computer Graphics Theory and Applications} (Lisbon, Portugal, 2 2023), GRAPP '23, SciTePress, pp.~64--74.

\bibitem[CKCS25]{chang2025design}
\textsc{Chang Z., Koulieris G.~A., Chang H.~J., Shum H.~P.}:
\newblock On the design fundamentals of diffusion models: A survey.
\newblock \emph{Pattern Recognition} (2025), 111934.

\bibitem[CPP{\etalchar{*}}]{cenready}
\textsc{Cen Z., Pi H., Peng S., Shuai Q., Shen Y., Bao H., Zhou X., Hu R.}:
\newblock Ready-to-react: Online reaction policy for two-character interaction generation.
\newblock In \emph{The Thirteenth International Conference on Learning Representations} (Singapore EXPO, Singapore).

\bibitem[CTO{\etalchar{*}}23]{chopin2023interaction}
\textsc{Chopin B., Tang H., Otberdout N., Daoudi M., Sebe N.}:
\newblock Interaction transformer for human reaction generation.
\newblock \emph{IEEE Transactions on Multimedia 25} (2023), 8842--8854.

\bibitem[CTR{\etalchar{*}}24]{cohan2024flexible}
\textsc{Cohan S., Tevet G., Reda D., Peng X.~B., van~de Panne M.}:
\newblock Flexible motion in-betweening with diffusion models.
\newblock In \emph{ACM SIGGRAPH 2024 Conference Papers} (Colorado Convention Center, Denver, CO, USA, 2024), ACM SIGGRAPH, pp.~1--9.

\bibitem[CWKS25]{chang2025large}
\textsc{Chang Z., Wang H., Koulieris G.~A., Shum H.~P.}:
\newblock Large-scale multi-character interaction synthesis.
\newblock \emph{arXiv preprint arXiv:2505.14087} (2025).

\bibitem[GDG{\etalchar{*}}24]{ghosh2024remos}
\textsc{Ghosh A., Dabral R., Golyanik V., Theobalt C., Slusallek P.}:
\newblock Remos: 3d motion-conditioned reaction synthesis for two-person interactions.
\newblock In \emph{European Conference on Computer Vision (ECCV)} (Milan, Italy, 2024), Springer.

\bibitem[GMJ{\etalchar{*}}24]{guo2024momask}
\textsc{Guo C., Mu Y., Javed M.~G., Wang S., Cheng L.}:
\newblock Momask: Generative masked modeling of 3d human motions.
\newblock In \emph{Proceedings of the IEEE/CVF Conference on Computer Vision and Pattern Recognition} (Seattle Convention Center, Seattle, WA, USA, 2024), IEEE/CVF, pp.~1900--1910.

\bibitem[HAB20]{henter2020moglow}
\textsc{Henter G.~E., Alexanderson S., Beskow J.}:
\newblock Moglow: Probabilistic and controllable motion synthesis using normalising flows.
\newblock \emph{ACM Transactions on Graphics (TOG) 39}, 6 (2020), 1--14.

\bibitem[HKPP20]{holden2020learned}
\textsc{Holden D., Kanoun O., Perepichka M., Popa T.}:
\newblock Learned motion matching.
\newblock \emph{ACM Transactions on Graphics (TOG) 39}, 4 (2020), 53--1.

\bibitem[HKS17]{holden2017phase}
\textsc{Holden D., Komura T., Saito J.}:
\newblock Phase-functioned neural networks for character control.
\newblock \emph{ACM Transactions on Graphics (TOG) 36}, 4 (2017), 1--13.

\bibitem[HRU{\etalchar{*}}17]{heusel2017gans}
\textsc{Heusel M., Ramsauer H., Unterthiner T., Nessler B., Hochreiter S.}:
\newblock Gans trained by a two time-scale update rule converge to a local nash equilibrium.
\newblock \emph{Advances in neural information processing systems 30} (2017).

\bibitem[HSK16]{holden2016deep}
\textsc{Holden D., Saito J., Komura T.}:
\newblock A deep learning framework for character motion synthesis and editing.
\newblock \emph{ACM Transactions on Graphics (TOG) 35}, 4 (2016), 1--11.

\bibitem[HSKJ15]{holden2015learning}
\textsc{Holden D., Saito J., Komura T., Joyce T.}:
\newblock Learning motion manifolds with convolutional autoencoders.
\newblock In \emph{SIGGRAPH Asia 2015 technical briefs}. ACM, 2015, pp.~1--4.

\bibitem[JGP17]{jang2017categorical}
\textsc{Jang E., Gu S., Poole B.}:
\newblock Categorical reparameterization with gumbel-softmax.
\newblock In \emph{International Conference on Learning Representations} (Toulon, France, 2017).

\bibitem[LH19]{loshchilov2018decoupled}
\textsc{Loshchilov I., Hutter F.}:
\newblock Decoupled weight decay regularization.
\newblock In \emph{International Conference on Learning Representations} (New Orleans, Louisiana, USA, 2019).

\bibitem[LSY{\etalchar{*}}23]{lee2023questenvsim}
\textsc{Lee S., Starke S., Ye Y., Won J., Winkler A.}:
\newblock Questenvsim: Environment-aware simulated motion tracking from sparse sensors.
\newblock In \emph{ACM SIGGRAPH 2023 Conference Proceedings} (Los Angeles Convention Center, Los Angeles, CA, USA, 2023), ACM SIGGRAPH, pp.~1--9.

\bibitem[LZL{\etalchar{*}}24]{liang2024intergen}
\textsc{Liang H., Zhang W., Li W., Yu J., Xu L.}:
\newblock Intergen: Diffusion-based multi-human motion generation under complex interactions.
\newblock \emph{International Journal of Computer Vision} (2024), 1--21.

\bibitem[{Mix}]{mixamo}
\textsc{{Mixamo}}:
\newblock Mixamo.
\newblock \url{https://www.mixamo.com/}.
\newblock Accessed: 2024-09-23.

\bibitem[MSHL22]{men2022gan}
\textsc{Men Q., Shum H.~P., Ho E.~S., Leung H.}:
\newblock Gan-based reactive motion synthesis with class-aware discriminators for human--human interaction.
\newblock \emph{Computers \& Graphics 102} (2022), 634--645.

\bibitem[MZ20]{gdcvault2024motionmatching}
\textsc{Mach M., Zhuravlov M.}:
\newblock Motion matching in the last of us part ii, 2020.
\newblock Accessed: 29-Sep-2024.

\bibitem[PMW23]{peng2023trajectory}
\textsc{Peng X., Mao S., Wu Z.}:
\newblock Trajectory-aware body interaction transformer for multi-person pose forecasting.
\newblock In \emph{Proceedings of the IEEE/CVF Conference on Computer Vision and Pattern Recognition} (Vancouver Convention Center, Vancouver, BC, Canada, 2023), IEEE/CVF, pp.~17121--17130.

\bibitem[PYAP22]{ponton2022combining}
\textsc{Ponton J., Yun H., Andujar C., Pelechano N.}:
\newblock Combining motion matching and orientation prediction to animate avatars for consumer-grade vr devices.

\bibitem[SGY{\etalchar{*}}24]{siyao2024duolando}
\textsc{Siyao L., Gu T., Yang Z., Lin Z., Liu Z., Ding H., Yang L., Loy C.~C.}:
\newblock Duolando: Follower gpt with off-policy reinforcement learning for dance accompaniment.
\newblock \emph{arXiv preprint arXiv:2403.18811} (2024).

\bibitem[SKY07]{shum07simulating}
\textsc{Shum H. P.~H., Komura T., Yamazaki S.}:
\newblock Simulating competitive interactions using singly captured motions.
\newblock In \emph{Proceedings of the 2007 ACM Symposium on Virtual Reality Software and Technology} (New York, NY, USA, 11 2007), VRST '07, ACM, pp.~65--72.

\bibitem[SKY10]{shum2010simulating}
\textsc{Shum H.~P., Komura T., Yamazaki S.}:
\newblock Simulating multiple character interactions with collaborative and adversarial goals.
\newblock \emph{IEEE Transactions on Visualization and Computer Graphics 18}, 5 (2010), 741--752.

\bibitem[SMK22]{starke2022deepphase}
\textsc{Starke S., Mason I., Komura T.}:
\newblock Deepphase: Periodic autoencoders for learning motion phase manifolds.
\newblock \emph{ACM Transactions on Graphics (TOG) 41}, 4 (2022), 1--13.

\bibitem[SSH{\etalchar{*}}24]{starke2024categorical}
\textsc{Starke S., Starke P., He N., Komura T., Ye Y.}:
\newblock Categorical codebook matching for embodied character controllers.
\newblock \emph{ACM Transactions on Graphics (TOG) 43}, 4 (2024), 1--14.

\bibitem[STKB24]{shafir2024human}
\textsc{Shafir Y., Tevet G., Kapon R., Bermano A.~H.}:
\newblock Human motion diffusion as a generative prior.
\newblock In \emph{The Twelfth International Conference on Learning Representations} (Vienna, Austria, 2024).

\bibitem[SZKZ20]{starke2020local}
\textsc{Starke S., Zhao Y., Komura T., Zaman K.}:
\newblock Local motion phases for learning multi-contact character movements.
\newblock \emph{ACM Transactions on Graphics (TOG) 39}, 4 (2020), 54--1.

\bibitem[TF23]{tanaka2023role}
\textsc{Tanaka M., Fujiwara K.}:
\newblock Role-aware interaction generation from textual description.
\newblock In \emph{Proceedings of the IEEE/CVF international conference on computer vision} (Paris, France, 2023), IEEE/CVF, pp.~15999--16009.

\bibitem[TZZ{\etalchar{*}}23]{tanke2023social}
\textsc{Tanke J., Zhang L., Zhao A., Tang C., Cai Y., Wang L., Wu P.-C., Gall J., Keskin C.}:
\newblock Social diffusion: Long-term multiple human motion anticipation.
\newblock In \emph{Proceedings of the IEEE/CVF International Conference on Computer Vision} (Paris, France, 2023), IEEE/CVF, pp.~9601--9611.

\bibitem[WXNW21]{wang2021multi}
\textsc{Wang J., Xu H., Narasimhan M., Wang X.}:
\newblock Multi-person 3d motion prediction with multi-range transformers.
\newblock \emph{Advances in Neural Information Processing Systems 34} (2021), 6036--6049.

\bibitem[XWG23]{xu2023stochastic}
\textsc{Xu S., Wang Y.-X., Gui L.}:
\newblock Stochastic multi-person 3d motion forecasting.
\newblock In \emph{The Eleventh International Conference on Learning Representations} (Kigali, Rwanda, 2023), ICLR.

\bibitem[XZY{\etalchar{*}}24]{xu2024regennet}
\textsc{Xu L., Zhou Y., Yan Y., Jin X., Zhu W., Rao F., Yang X., Zeng W.}:
\newblock Regennet: Towards human action-reaction synthesis.
\newblock In \emph{Proceedings of the IEEE/CVF Conference on Computer Vision and Pattern Recognition} (Seattle Convention Center, Seattle, WA, USA, 2024), IEEE/CVF, pp.~1759--1769.

\bibitem[YKJ{\etalchar{*}}19]{yoon2019robots}
\textsc{Yoon Y., Ko W.-R., Jang M., Lee J., Kim J., Lee G.}:
\newblock Robots learn social skills: End-to-end learning of co-speech gesture generation for humanoid robots.
\newblock In \emph{2019 International Conference on Robotics and Automation (ICRA)} (Montreal, QC, Canada, 2019), IEEE, IEEE, pp.~4303--4309.

\bibitem[YZS{\etalchar{*}}23]{yang2023diffusurvey}
\textsc{Yang L., Zhang Z., Song Y., Hong S., Xu R., Zhao Y., Zhang W., Cui B., Yang M.-H.}:
\newblock Diffusion models: A comprehensive survey of methods and applications.
\newblock \emph{ACM Computing Surveys 56}, 4 (2023), 1--39.

\bibitem[ZSKS18]{zhang2018mode}
\textsc{Zhang H., Starke S., Komura T., Saito J.}:
\newblock Mode-adaptive neural networks for quadruped motion control.
\newblock \emph{ACM Transactions on Graphics (TOG) 37}, 4 (2018), 1--11.

\end{thebibliography}


\end{document}